\documentclass[twocolumn,prc,showpacs,superscriptaddress,preprintnumbers,amsmath,amssymb]{revtex4-1}
\usepackage{dcolumn}
\usepackage{bm}
\usepackage{graphicx}

\usepackage{color}
\definecolor{dgreen}{cmyk}{1.,0.,1.,0.4}        
\definecolor{orange}{cmyk}{0.,0.353,1.,0.}    

\begin{document}
\title{Event-plane flow analysis without non-flow effects}
\author{Ante Bilandzic}
\author{Naomi van der Kolk}
\affiliation{Nikhef, Science Park 105, 1098 XG Amsterdam,The Netherlands}
\affiliation{Utrecht University, P.O. Box 80000, 3508 TA Utrecht, 
The Netherlands}
\author{Jean-Yves~Ollitrault} 
\affiliation{Institut de Physique Th{\'e}orique, CEA-Saclay,
F-91191 Gif-sur-Yvette cedex}
\author{Raimond Snellings}
\affiliation{Utrecht University, P.O. Box 80000, 3508 TA Utrecht, 
The Netherlands}
\date{\today}
\begin{abstract}
The event-plane method, which is widely used to analyze anisotropic flow
in nucleus-nucleus collisions, is 
known to be biased by nonflow effects, especially at high $p_t$. 
Various methods (cumulants, Lee-Yang zeroes) have been proposed to 
eliminate nonflow effects, but their implementation is tedious, which 
has limited their application so far. In this paper, we show that the 
Lee-Yang-zeroes method can be recast in a form similar to the standard 
event-plane analysis. Nonflow correlations are strongly suppressed by using the 
information from the length of the flow vector, in addition to the event-plane angle. 
This opens the way to improved analyses of elliptic 
flow and azimuthally-sensitive observables at RHIC and LHC. 
\end{abstract}

\maketitle

\section{Introduction}

Studies of particle production at the BNL Relativistic Heavy Ion
Collider (RHIC) have revealed strong collective effects: in
particular, the azimuthal distribution transverse to the direction of
the colliding nuclei has sizable anisotropies, a phenomenon called 
anisotropic flow. The main component of this anisotropy, elliptic
flow, has been extensively measured for several beam energies and 
collision systems~\cite{Ackermann:2000tr,Back:2002gz,Adler:2003kt}. 

Anisotropic flow is most often analyzed using the event-plane 
method~\cite{Poskanzer:1998yz}. This analysis technique is plagued by 
systematic errors due to nonflow effects~\cite{Ollitrault:1995dy}. 
There are other sources of systematic errors, such as 
fluctuations~\cite{Miller:2003kd,Alver:2006wh}, but 
nonflow effects are expected to be the dominant source of error at 
high $p_t$~\cite{Adams:2004bi}, where they are likely to originate 
from jet-like (hard) correlations; they are expected to be even larger at
the LHC. 
The purpose of this paper is to show that nonflow effects can be
suppressed at the expense of a slight modification of the event-plane method. 

Anisotropic flow of selected produced particles, in a given 
part of phase-space, is defined as their azimuthal 
correlation with the reaction plane~\cite{Voloshin:1994mz}
\begin{equation}
\label{defvn}
v_n\equiv \left\langle\cos(n(\phi-\Phi_{RP}))\right\rangle
\end{equation}
where $n$ is an integer ($v_1$ is {\it directed\/} flow, $v_2$ is {\it
elliptic\/} flow), $\phi$, $\Phi_{RP}$ and angular brackets denote
respectively the azimuth of the particle under study, the azimuth 
of the reaction plane, and an average over particles 
and events. Since $\Phi_{RP}$ is not known experimentally, $v_n$ cannot be 
measured directly. 

The most commonly used method to estimate $v_n$ is 
the event-plane method~\cite{Poskanzer:1998yz}.
In each event, one constructs an estimate of the reaction plane 
$\Phi_{RP}$, the ``event plane'' $\Phi_{EP}$~\cite{Danielewicz:1985hn}. 
The anisotropic flow coefficients 
are then estimated as 
\begin{equation}
\label{v2std}
v_n{\{\rm EP}\}\equiv \frac{1}{R} \left\langle\cos(n(\phi-\Phi_{EP}))\right\rangle,
\end{equation}
where $R=\left\langle\cos(n(\Phi_{EP}-\Phi_{RP}))\right\rangle$ is 
the event-plane resolution, which corrects for the difference
between $\Phi_{EP}$ and $\Phi_{RP}$. This resolution 
is determined in each class of events through a standard 
procedure~\cite{Ollitrault:1997di}.

The analogy between Eq.~(\ref{v2std}) and Eq.~(\ref{defvn}) makes the method 
rather intuitive, but its practical implementation has a few subtleties:
\begin{itemize}
\item{One must remove autocorrelations: the particle under study
should not be used in defining the event plane, otherwise there is a
trivial correlation between $\phi$ and
$\Phi_{EP}$~\cite{Danielewicz:1985hn}. This means in practice that 
one must keep track of which particles have been used in defining the event plane,
so as to remove them if necessary.}
\item{More generally, there are sources of correlation, other than
flow, through which the particle under study can be correlated with 
a particle used in defining the event plane. Such correlations,
called ``nonflow effects'',   result in $v_n\{{\rm EP}\}\not= v_n$
and must be suppressed. This cannot be done in a 
systematic way, but rapidity gaps are believed to reduce 
nonflow effects~\cite{Adler:2003kt,Voloshin:2006wi}. }
\item{Event-plane flattening procedures must be implemented to correct
for azimuthal asymmetries of the detector
acceptance~\cite{Poskanzer:1998yz}.} 
\end{itemize}

A systematic way of suppressing nonflow effects is to use improved
methods such as  cumulants~\cite{Borghini:2001vi} or Lee-Yang 
zeroes~\cite{Bhalerao:2003xf}. Cumulants have been used at 
SPS~\cite{Alt:2003ab} and RHIC~\cite{Adler:2002pu,Adams:2004bi}.
Lee-Yang zeroes have been implemented at SIS~\cite{Bastid:2005ct} and 
at RHIC~\cite{Abelev:2008ed}. 
They are comparatively much less used than the event-plane method, and
one reason is that the event-plane method is deemed more intuitive and
handy. 

In this paper, we show that the method of flow analysis
based on Lee-Yang zeroes can be rewritten in a way which is
mathematically equivalent to the original
formulation~\cite{Bhalerao:2003xf}, but formally analogous to the
event-plane method, which makes it more intuitive. 
The corresponding estimate of $v_n$ is defined as 
\begin{equation}
\label{v2LYZ}
v_n\{{\rm LYZ}\}\equiv \left\langle W_R\cos(n(\phi-\Phi_{EP}))\right\rangle,
\end{equation}
where $\Phi_{EP}$ is the same as in Eq.~(\ref{v2std}), and
$W_R$ is an event weight as defined in this paper. 
The formal analogy with the event-plane method, Eq.~(\ref{v2std}), is
obvious. 
The advantage of the improved event-plane method defined by
Eq.~(\ref{v2LYZ}) over the standard 
event-plane method is that {\it both autocorrelations and nonflow effects\/} 
are automatically suppressed. 

The paper is organized as follows. In Sec.~\ref{s:description}, 
we describe the method for a detector with perfect azimuthal symmetry, 
and we explain why it automatically removes autocorrelations 
and nonflow correlations, in contrast to the standard event-plane
method. 
Readers interested in applying the method should read 
Appendix~\ref{s:implementation}, which describes the 
recommended practical implementation, taking into account 
anisotropies in the detector acceptance.
In Sec.~\ref{s:results}, we present results of Monte-Carlo
simulations, where results obtained with the Lee-Yang zeroes method
are compared to those obtained with 2- and 4-particle cumulants. 
Sec.~\ref{s:discussion} concludes with a discussion of where 
the method should be applicable, and of its limitations. 

\section{Description of the method}
\label{s:description}

\subsection{The flow vector}
\label{s:flowvector}

The first step of the flow analysis is to evaluate, for each
event, the flow vector of the event.
It is a two-dimensional vector ${\bf Q}=(Q_x,Q_y)$ defined as
\begin{eqnarray}
\label{flowvector}
Q_x= Q\cos(n\Phi_{EP})&\equiv&\sum_{j=1}^M w_j\,\cos(n\phi_j)\cr
Q_y= Q\sin(n\Phi_{EP})&\equiv&\sum_{j=1}^M w_j\,\sin(n\phi_j),
\end{eqnarray}
where the sum runs over all detected particles~\cite{footnote1}. 
$M$ is the observed multiplicity of the event,
$\phi_j$ are the azimuthal angles of the particles measured with
respect to a fixed direction in the laboratory.
The coefficients $w_j$ in Eq.~(\ref{flowvector}) are weights depending
on transverse momentum, particle mass and rapidity.
The best weight, which minimizes the statistical error 
(or, equivalently, maximizes the resolution) is $v_n$ itself,
$w_j(p_T,y)\propto v_n(p_T,y)$~\cite{Borghini:2000sa}.
A reasonable choice for elliptic flow measurements at RHIC (and probably LHC) 
is $w=p_T$. 

\begin{figure}
\centerline{\includegraphics*[width=1.\linewidth]{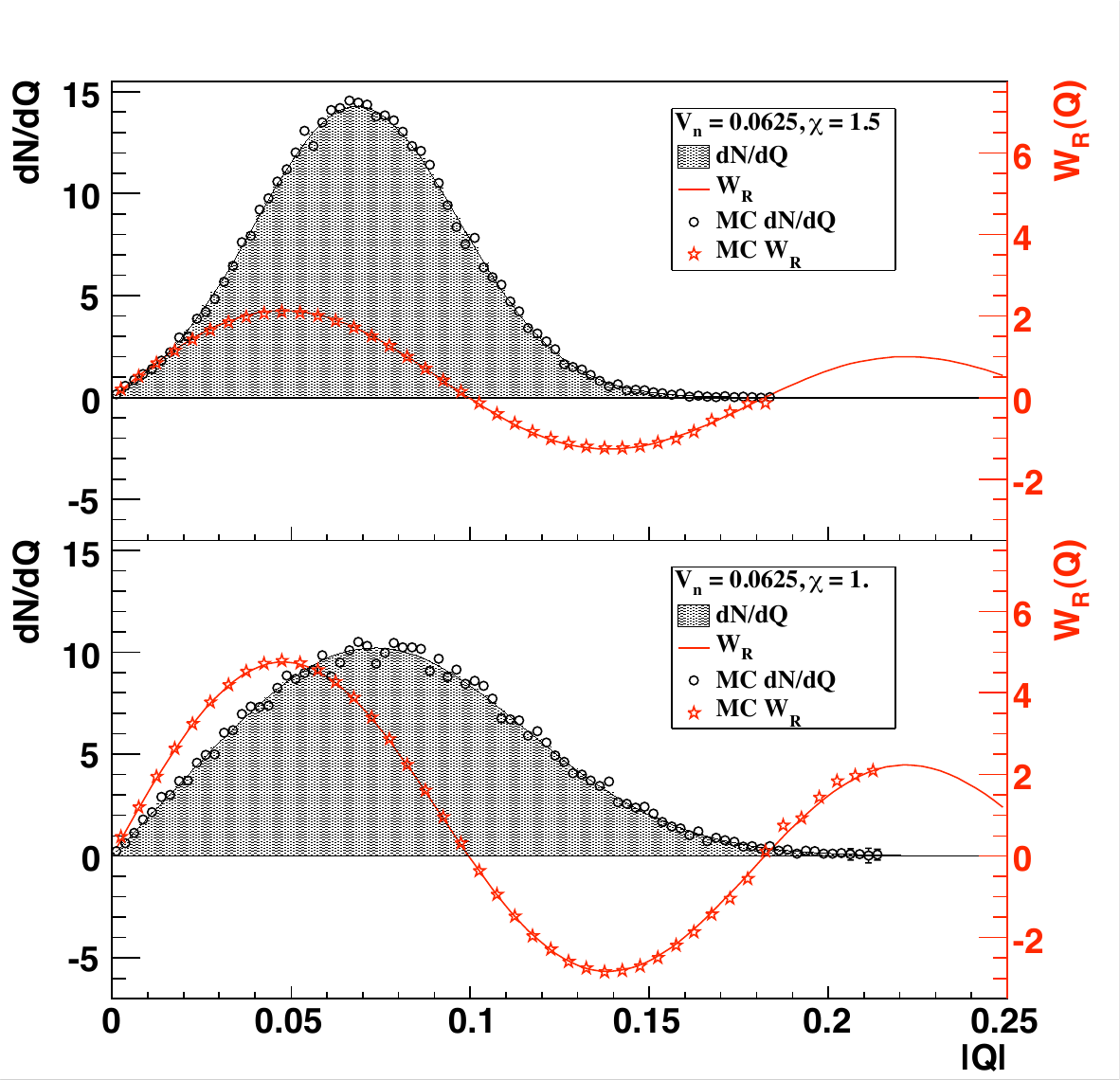}}
\caption{(Color online) 
Shaded area: probability distribution of $Q$, Eq.~(\ref{qdist}), with
$V_n=0.0625$. 
Top: $\chi=1.5$, corresponding to a resolution $R=0.86$ in the
standard analysis  (see Eq.~(\ref{v2std})). Bottom: $\chi=1$,
corresponding to a resolution $R=0.71$.  This is the typical value for
a semi-central Au-Au collision at RHIC analyzed by  the STAR
TPC~\cite{Adams:2004bi}.
Solid curve: weight $W_R$ defined by Eqs.~(\ref{wsimple}) and
(\ref{defC}). 
Open circles: histograms of
the distribution of $Q$ obtained in the Monte-Carlo simulation of
Sec.~\ref{s:results}, following the procedure detailed in
Appendix~\ref{s:implementation}.  
Stars: weights obtained in Sec.~\ref{s:results}.
} 
\label{fig:weight}
\end{figure}

If collective flow is present, the azimuthal angles $\phi_j$ and the
event plane $\Phi_{EP}$ are correlated with the true reaction plane
$\Phi_{RP}$, and the goal of the flow analysis is to measure this
correlation. This is usually done within a set of events belonging to
the same centrality class. Integrated flow is defined as the average
value of the  projection of ${\bf Q}$ onto 
the true reaction plane:
\begin{equation}
\label{integrated}
V_n\equiv \left\langle Q\cos(n(\Phi_{EP}-\Phi_{RP}))\right\rangle
\end{equation}
where angular brackets denote an average over events in the same 
centrality class. 
We use a capital letter for $V_n$ because it is in general a
dimensionful quantity: it is the weighted sum of the $v_n$'s of
individual particles, according to Eqs.~(\ref{defvn}) and
(\ref{flowvector}). 
The flow vector fluctuates around this average value because the
multiplicity is finite. 
These fluctuations can be modeled using the central limit theorem. 
The resulting distribution of $Q$ is~\cite{Ollitrault:1995dy}:
\begin{equation}
\label{qdist}
\frac{dN}{dQ}=\frac{2\chi^2 Q}{V_n^2}\exp\left(-\chi^2\left(\frac{Q^2}{V_n^2} + 1\right)\right)
I_0\left(\frac{2\chi^2 Q}{V_n}\right),
\end{equation}
where $\chi$ is a dimensionless quantity called the resolution
parameter, which characterizes the relative magnitude of collective
flow and statistical fluctuations. 
The resolution $R$ in Eq.~(\ref{v2std}) increases from $0$ to $1$ as
$\chi$ goes from $0$ to $+\infty$. 
Fig.~\ref{fig:weight} illustrates the distribution of $Q$ for two values
of $\chi$. For $\chi\gg 1$, this distribution is a narrow peak centered at
$Q=V_n$.

Lee-Yang zeroes use the projection of 
the flow vector onto a fixed,
arbitrary direction making an angle $n\theta$ with respect
to the $x$-axis. We denote this projection by
$Q_\theta$: 
\begin{equation}
\label{defqtheta}
Q_\theta \equiv Q_x\cos n\theta+Q_y\sin n\theta
=Q\cos(n(\Phi_{EP}-\theta)) .
\end{equation}

\subsection{Integrated flow}
\label{s:integrated}

We now explain how 
the integrated flow $V_n$, defined by Eq.~(\ref{integrated}), is obtained. 
We define the complex-valued function:
\begin{equation}
\label{genx}
G_\theta(r)\equiv\left\langle  e^{i r
Q_\theta}\right\rangle\equiv\frac{1}{N_{\rm evts}}\sum_{\rm events} e^{i r
Q_\theta}.
\end{equation}
If there is no collective flow, the probability distribution of
$Q_\theta$ is a Gaussian due to the central limit theorem (if $M\gg
1$). Its Fourier transform $G_\theta(r)$ is also a Gaussian. 
Collective flow results in oscillations of $G_\theta(r)$ around zero: 
In the ideal case where the multiplicity is so large that
 fluctuations can be neglected, $\Phi_{EP}\simeq\Phi_{RP}$ and
 $Q\simeq V_n$. Inserting Eq.~(\ref{defqtheta}) into (\ref{genx}) and 
averaging over $\Phi_{RP}$, one obtains 
\begin{equation}
\label{genideal}
G_\theta(r)\simeq J_0(rV_n),
\end{equation}
where $J_0(x)$ denotes the Bessel function of the first kind of order
0, which oscillates around 0. Finite multiplicity fluctuations result
in a gaussian smearing of $G_\theta(r)$, but quite remarkably, the
location of the zeroes is unchanged, up to statistical fluctuations due
to the finite number of events~\cite{Bhalerao:2003yq}.

As a consequence, 
the modulus $|G_\theta(r)|$ has sharp minima for positive $r$, which
can be estimated numerically. 
The position of the first minimum, $r_\theta$, 
is used to estimate $V_n$, 
using Eq.~(\ref{genideal}):
\begin{equation} 
\label{vnzero}
V_n=\frac{j_{01}}{r_\theta},
\end{equation}
where $j_{01}\simeq 2.40483$ is the first zero of $J_0(x)$. 
One may also check, as a consistency test, that
$|G_\theta(r_\theta)|=0$ within statistical errors~\cite{Bhalerao:2003yq}.

The above procedure only makes use of the projection 
of the flow vector onto an arbitrary direction $\theta$. 
For a perfect detector, azimuthal symmetry ensures that  
$r_\theta$ is independent of $\theta$, up to statistical errors. 
In practice, however, it is recommended to repeat the analysis 
for several values of $\theta$ 
(see Appendix~\ref{s:implementation}).

\subsection{Differential flow and event weight}

We now derive the expression of the event weight in Eq.~(\ref{v2LYZ}),
which is the crucial improvement of our paper over the standard
event-plane method. 
The goal is to measure the differential flow $v_n$ of selected
produced particles. 
$v_n$ can be obtained by shifting the weights $w_j$ of the 
selected particles in Eq.~(\ref{flowvector}) by an infinitesimal
quantity $\varepsilon$, $w'_j=w_j+\varepsilon$, and computing the
integrated flow $V'_n$ with the new weights. The differential flow is
then simply given by $v_n=\delta V_n/\varepsilon$, with $\delta
V_n=V'_n-V_n$. Differentiating  Eq.~(\ref{vnzero}), 
\begin{equation}
\label{diffvn0}
v_n\{\rm LYZ\}=\frac{\delta
 V_n}{\varepsilon}=-\frac{V_n}{\varepsilon}\frac{\delta
 r_\theta}{r_\theta}, 
\end{equation}
where $\delta r_\theta$ denotes the shift of the zero. 
Differentiating the condition $\langle e^{ir_\theta
 Q_\theta}\rangle=0$, one obtains
\begin{equation}
\label{dgenx}
\delta r_\theta\left\langle Q_\theta  e^{i r_\theta Q_\theta}\right\rangle+
r_\theta\left\langle \delta Q_\theta  e^{i r_\theta Q_\theta}\right\rangle=0.
\end{equation}
For an event containing one selected particle, 
Eqs.~(\ref{flowvector}) and (\ref{defqtheta}) give 
$\delta Q_\theta=\varepsilon\cos(n(\phi-\theta))$, where $\phi$ is the
azimuth of the selected particle. Eq.~(\ref{diffvn0}) then gives
\begin{equation}
\label{diffvn1}
v_n\{\rm LYZ\}=V_n\frac{\left\langle\cos(n(\phi-\theta))e^{i r_\theta Q_\theta}
 \right\rangle}{\left\langle  Q_\theta  e^{i r_\theta
     Q_\theta}\right\rangle}, 
\end{equation}
where the average in the numerator is over selected particles,
and the average in the denominator is over events. 
In this expression, $\theta$ is an arbitrary reference angle.
Both the numerator and the denominator are expected to be independent
of $\theta$, up to asymmetries in the detector acceptance, and
statistical fluctuations. 
In practice, we recommend  to first take the ratio and then average 
over $\theta$, as explained in Sec.\ref{s:wandpsi}. 
Here, we derive simple approximate expressions by assuming that 
$r_\theta$ is independent of $\theta$, and by averaging 
the numerator and the denominator over $\theta$ {\it
 before\/} taking the ratio. We thus obtain: 
 \begin{equation}
\label{J1}
v_n\{\rm LYZ\}=V_n\frac{\left\langle\cos(n(\phi-\Phi_{EP}))J_1(r_\theta Q)
 \right\rangle}{\left\langle  Q J_1(r_\theta Q)\right\rangle},
\end{equation}
where $J_1(x)$ is the derivative of $-J_0(x)$. 
Identifying Eq.~(\ref{J1}) with Eq.~(\ref{v2LYZ}), we obtain the event
weight  
\begin{equation}
\label{wsimple}
W_R\equiv \frac{1}{C}J_1(r_\theta Q),
\end{equation}
where $C$ is a normalization constant which can be computed using 
the distribution (\ref{qdist}): 
\begin{equation}
\label{defC}
C=\frac{1}{V_n}\left\langle  Q J_1(r_\theta Q)\right\rangle
=\exp\left(-\frac{j_{01}^2}{4\chi^2}\right) J_1(j_{01}).
\end{equation}
The difference with the standard event-plane analysis is that each event
is given a weight (\ref{wsimple}) which depends on 
the length of the flow vector $Q$, a quantity which is not used in the
standard analysis.  
Eq.~(\ref{wsimple}) involves the integrated flow $V_n$ through
$r_\theta$, which must  be determined in a first pass through the
data. 

Fig.~\ref{fig:weight} displays the variation of
$W_R$ with $Q$, for two values of the resolution parameter. 
For $\chi\gg 1$, the distribution of $Q$
is a narrow peak centered at $Q=V_n$. 
Therefore, the weight defined by Eqs.~(\ref{wsimple}) and (\ref{defC}) is 
close to $1$ for all events.
If $\chi$ is smaller, the distribution of $Q$ is 
wider, and $W_R$ is negative for some events. 
These negative weights are required in order to subtract nonflow
effects. 
On the other hand, they also subtract part of the flow. 
In order to compensate for this effect, the global normalization of the
weight increases when $\chi$ decreases 
(as illustrated in Fig.~\ref{fig:weight} by the fact that the amplitude of the curve 
showing the weight changes for different values of $\chi$).
This qualitatively
explains the $\chi$ dependence in Eq.~(\ref{defC}). 

The weight (\ref{wsimple}) vanishes linearly at $Q=0$. 
This is physically intuitive. 
Given that the flow
vector is obtained by summing over all particles, 
one increases the relative weight of
collective flow over individual, random motion of the particles. 
If the flow vector is small in an event, it means that the random 
motion hides the collective motion in this particular event, which is 
therefore of little use for the flow analysis.

\subsection{Nonflow effects and autocorrelations}
\label{s:nonflow}

We now explain why the method suppresses
nonflow effects and autocorrelations  
on the basis of two simple examples.

As a first example, we assume that each particle splits into two
particles with identical momenta, roughly imitating the effect of
resonance decays or track splitting in a detector. 
This splitting does not change the anisotropic flow $v_n$, defined by
Eq.~(\ref{defvn}), but it introduces 
nonflow correlations, which bias standard analyses as will be shown in 
Sec.~\ref{s:results}. 
The splitting leaves $v_n\{\rm LYZ\}$ unchanged:
it multiples both the flow vector, Eq.~(\ref{flowvector})
and the integrated flow $V_n$, Eq.~(\ref{integrated}) by 2. Therefore
$r_\theta$ in Eq.~(\ref{vnzero}) is divided by 2, and $v_n\{\rm LYZ\}$
defined by Eq.~(\ref{diffvn1}) is unchanged. 

As a second example,
we consider the situation where there is collective flow in the 
system, but the selected
particles have $v_n=0$. 
We further assume that the selected particles are uncorrelated
 with the other particles. In the standard event-plane method, one
 needs to subtract the selected particles from the flow vector
 (\ref{flowvector}), otherwise autocorrelations yield
 $v_n\{EP\}>0$. We now show that $v_n\{LYZ\}=0$, even if selected
 particles are included in the flow vector. 

We separate the flow vector, Eq.~(\ref{flowvector}), into 
the contribution of selected particles, ${\bf Q}_{\rm sel.}$, 
and other particles ${\bf Q}_{\rm others}$.
\begin{equation}
\label{flownonflow}
{\bf Q}={\bf Q}_{\rm sel.}+{\bf Q}_{\rm others}.
\end{equation} 
Our estimate of $v_n$ is defined by Eq.(\ref{diffvn1}).
Since the flow vector appears in an exponential, 
the contributions of selected particles and other particles 
can be written as a product of two independent factors:
\begin{equation}
\label{separationfnf}
v_n=V_n\frac{
\left\langle\cos(n(\phi-\theta))e^{i r_\theta (Q_{\rm sel.})_\theta}
 \right\rangle
\left\langle e^{i r_\theta (Q_{\rm others})_\theta}
 \right\rangle
}{\left\langle  Q_\theta  e^{i r_\theta
     Q_\theta}\right\rangle}, 
\end{equation}

Let us define $G_{\rm others, \theta}(r)$ by replacing $Q_\theta$ with
$Q_{\rm   others,\theta}$ in Eq.~(\ref{genx}). 
Following the same reasoning as in Sec.~\ref{s:integrated}, 
the first zero of $G_{\rm others, \theta}$ depends on the integrated
flow $V_{n,{\rm others}}$ of other particles. 
We have assumed that $v_n=0$ for selected particles, therefore
$V_{n,{\rm others}}=V_n$, and 
\begin{equation}
\left\langle e^{i r_\theta (Q_{\rm others})_\theta}
 \right\rangle=
\left\langle e^{i r_\theta Q_\theta}
 \right\rangle=0.
\end{equation}
Inserting into Eq.~(\ref{separationfnf}), we find 
\begin{equation}
v_n\{{\rm LYZ}\}= 0,
\end{equation}
up to statistical fluctuations. 
This proof can easily be generalized to the situation where each selected
particle is correlated with a few additional particles (e.g. within a jet) 
which are not correlated with the bulk of particles producing
collective flow. 

We have constructed two simple examples where Lee-Yang zeroes are 
 able to eliminate nonflow effects and autocorrelations. In actual 
 experiments, however, flow and nonflow effects are likely to be 
 mingled, and detailed simulations must be carried out to determine 
 to what extent the suppression is effective.

\section{Simulations}
\label{s:results}

To check the validity of the procedure described in this paper and to compare it with other analysis methods 
$N=$ 28$\;$000 events were simulated with a Monte-Carlo program dubbed
GeVSim~\cite{gevsim}.  
In GeVSim the $v_2$ and the particle yield as function of transverse momentum and pseudorapidity
are generated according to a user-defined parameterization. 
For these simulations events were generated using a linear dependence of $v_2(p_t)$ 
in the range 0--2 GeV/$c$, above 2 GeV/$c$ the $v_2(p_t)$ was set constant.
The average elliptic flow is $\langle v_2\rangle =0.0625$. 
We then reconstructed $v_2(p_t)$ from the simulated events
using several methods: the Lee-Yang-zeroes method described in
Appendix~\ref{s:implementation}, as well as 2- and 4-particle
cumulants~\cite{Borghini:2001vi}. The corresponding estimates of $v_2$
are denoted by $v_2\{{\rm LYZ}\}$, $v_2\{2\}$ and $v_2\{4\}$,
respectively. $v_2\{2\}$ is generally close to $v_2$ from the
traditional event-plane method; both are biased by nonflow effects. 
On the other hand, $v_2\{4\}$ is expected to be close to $v_2\{{\rm
LYZ}\}$, with the bias from nonflow effects suppressed. 
The weight $w_j$ in Eq.~(\ref{flowvector}) was chosen identically
$1/M$ for all particles, with $M$ the event multiplicity, so that the
integrated flow $V_n$ defined by Eq.~(\ref{defvn}) coincides with the 
average elliptic flow, i.e., $V_n=0.0625$. 
The analysis was repeated twice by varying the multiplicity $M$
used in the flow analysis: 
the values 256 and 576 were used, so as to achieve a resolution
of $\chi =$1 and 1.5.~\cite{footnote2}.

\begin{figure}
\centerline{\includegraphics*[width=1.\linewidth]{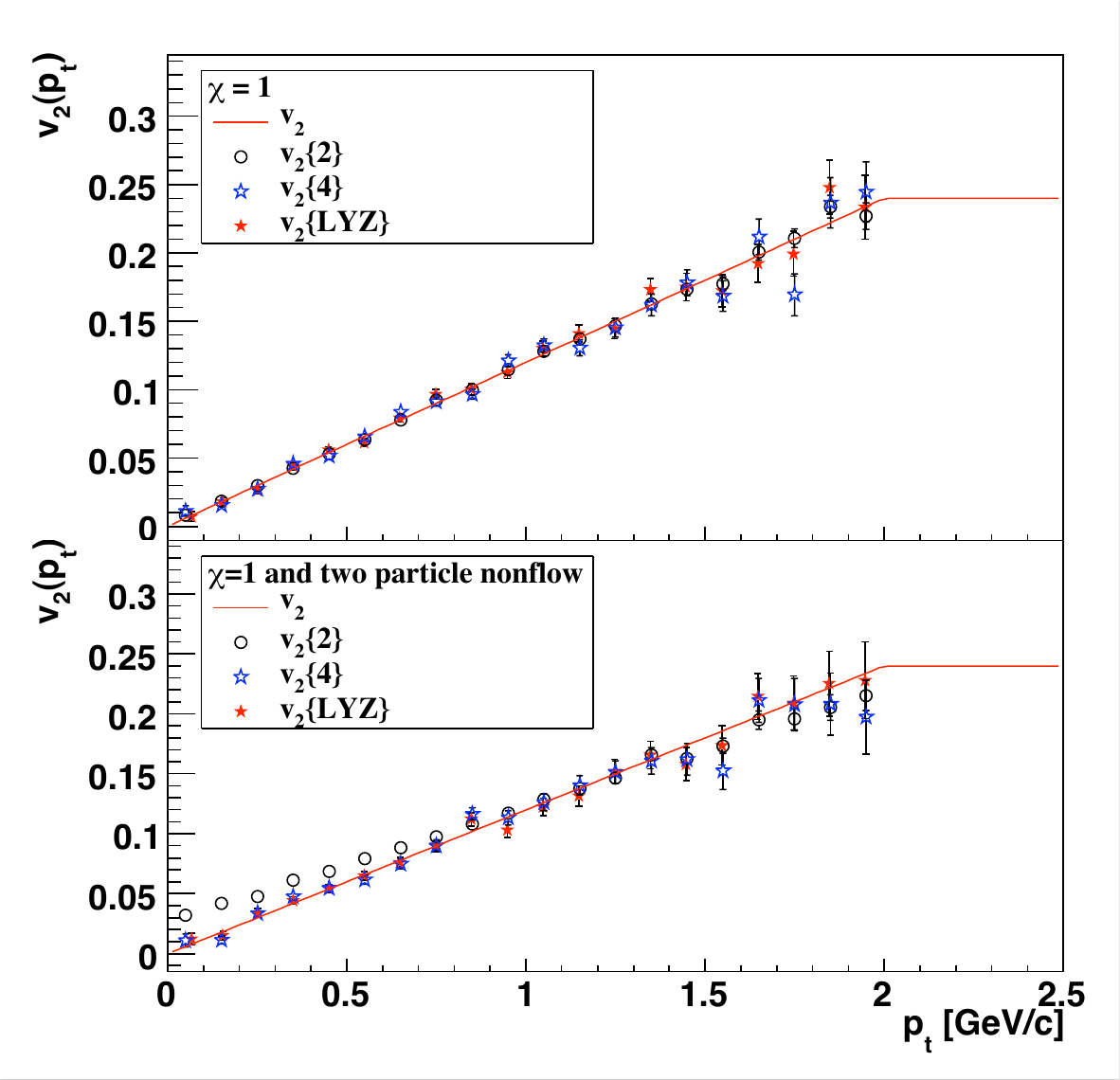}}
\caption{
(Color online)
Differential elliptic flow $v_2(p_t)$ reconstructed using different methods; in the upper panel from events where no nonflow was included, in the lower panel from events with nonflow.
 The line in both panels is the input $v_2$. 
}
\label{fig:simulation}
\end{figure}

Fig.~\ref{fig:simulation} shows the generated (input) $v_2(p_t)$
together with the reconstructed $v_2(p_t)$ using cumulants and
Lee-Yang zeroes for $\chi=1$.
The upper panel shows the results in the case where all correlations
are due to flow. In this case, all three methods yield the correct 
$v_2(p_t)$ and $\langle v_2\rangle$ 
within statistical uncertainties (see Table~\ref{Vint}),
which are twice larger for $v_2\{4\}$ and
$v_2\{{\rm LYZ}\}$ than for $v_2\{2\}$ (see Sec.~\ref{s:stat}). 

In the lower panel, simulations are shown which include nonflow
effects. Because GeVSim generates no nonflow, nonflow correlations 
are introduced by using each input track
twice, as in Sec.~\ref{s:nonflow}. 
Experiments at RHIC have shown~\cite{Adams:2004bi} that 
nonflow effects are larger at high-$p_t$ (probably due to jet-like
correlations),  and a realistic simulation of nonflow effects should
take into account this $p_t$ dependence. Our simplified
implementation,  which does not, is not realistic. It is merely an
illustration of the impact of nonflow effects on the flow analysis. 
Fig.~\ref{fig:simulation} shows that due to nonflow effects, the 
method based on two-particle cumulants ($v_2\{2\}$) overestimates the
average elliptic flow $\langle v_2\rangle$. 
The error on the average elliptic flow is larger than 
20\% (see Table~\ref{Vint}, right column). 
The transverse-momentum dependence of $v_2(p_t)$ is also not correct,
with an excess at low $p_t$ by $0.03$. 
By contrast, the results from 4-particle cumulants ($v_2\{4\}$) and
Lee-Yang zeroes ($v_2\{{\rm LYZ}\}$) are, within statistical
uncertainties, in agreement with the true generated flow distribution. 
This shows that the method presented in this paper is able to remove
nonflow effects.

\begin{table}[ht]
\caption{Value of the average elliptic flow $\langle v_2\rangle$
reconstructed, using different methods, from simulated data with and without nonflow effects. The input value is $\langle v_2\rangle
=0.0625$.} 
\label{Vint}
\smallskip
\begin{center}
\begin{tabular}{|c|c|c|}
\hline
Method& {\rm Flow only}&{\rm Flow+nonflow} \\ 
\hline
$v_2\{2\}$ &$ 0.0626\pm 0.0003$& $0.0764\pm 0.0004$\\
\hline
$v_2\{4\}$ & $0.0624\pm 0.0005$& $0.0627\pm 0.0007$\\
\hline
$v_2\{{\rm LYZ}\}$ & $0.0626\pm 0.0005$& $0.0629\pm 0.0007$\\
\hline
\end{tabular}
\end{center}
\end{table}

\section{Discussion}
\label{s:discussion}

Two effects limit the accuracy of flow analyses at high energy: 
nonflow effects and eccentricity
fluctuations~\cite{Miller:2003kd,Alver:2006wh}. 
The method presented in this paper is an improved event-plane method,
which strongly suppresses the first source of uncertainty, nonflow effects. 
It has been argued~\cite{Voloshin:2007pc} that cumulants 
(and therefore Lee-Yang zeroes, which corresponds to the limit of
large-order cumulants) also eliminate eccentricity
fluctuations~\cite{Miller:2003kd,Alver:2006wh}.
However, a detailed study~\cite{Alver:2008zza} shows that even
with cumulants, there may remain large effects of fluctuations in
central collisions and/or small systems. This issue deserves more
detailed investigation. 

Letting aside the question of fluctuations, we now discuss which
method of flow analysis should be used, depending on the situation. 
There are three main classes of methods: the standard 
event-plane method~\cite{Poskanzer:1998yz}, 
four-particle cumulants~\cite{Borghini:2001vi}, and the
Lee-Yang-zeroes method presented in this paper.  
When the standard event-plane method is used, 
nonflow effects and eccentricity fluctuations are generally the main
sources of uncertainty on $v_n$, and they dominate over statistical
errors. The magnitude of this uncertainty is 
at least 10\% at RHIC in semi-central collisions; it is larger for
more central or more peripheral collisions, and also larger at high
$p_t$. 
Unless statistical errors are of comparable magnitude as errors from
nonflow effects, cumulants or Lee-Yang zeroes should be preferred over
the standard method. 

The main advantage of Lee-Yang zeroes, compared to cumulants, 
is that the method involves an event-plane angle. This is useful in
particular for studying azimuthally dependent
correlations~\cite{Adams:2004wz,Bielcikova:2003ku}. 
Such studies cannot be done with cumulants, but they are
straightforward with Lee-Yang zeroes. 
The only complication is that the azimuthal distribution of particle
pairs generally involves {\it sine\/} terms~\cite{Borghini:2004ra}, in 
addition to the {\it cosine\/} terms of Eq.~(\ref{defvn}). 
These terms are simply obtained by replacing cos with sin in 
Eq.~(\ref{v2LYZ}). 

When studying anisotropic flow of individual particles, both cumulants
and Lee-Yang zeroes can be applied. 
The cumulant method has been recently improved by directly calculating
the cumulants~\cite{Bilandzic:2010jr}. With these improvements, both
methods are expected to be essentially equivalent. 
The slight advantages of Lee-Yang zeroes are: 
1) They are easier to implement. 
2) They further reduce the error from nonflow effects. 
3) The statistical error is slightly smaller if the resolution
parameter $\chi>1$. For $\chi=0.8$, the error is only 35\% larger with
Lee-Yang zeroes than with 4-particle cumulants (and 4 times larger
than with the event-plane method).

Our recommendation is that Lee-Yang zeroes should be
used as soon as $\chi>0.8$. For small values of $\chi$, 
typically $\chi<0.6$, statistical errors on Lee-Yang zeroes blow up
exponentially, which rules out the method; the statistical error on
4-particle cumulants also increases but more mildly, and their
validity extends down to lower values of the resolution if very large
event statistics is available. 

A limitation of the present method is that it does not 
apply to mixed harmonics: this means that it cannot be used to measure
$v_1$ and $v_4$ at RHIC and LHC using the event plane from elliptic 
flow~\cite{Adams:2003zg}. 
Note that $v_1$ can in principle be measured using Lee-Yang 
zeroes~\cite{Borghini:2004ad} using the ``product'' generating
function, but this method cannot be recast in the form of an improved
event-plane method. 
Higher harmonics such as $v_4$ also have a sensitivity to
autocorrelations and nonflow effects, which is significantly reduced by
using the product generating function~\cite{Borghini:2001vi}. 

In conclusion, we have presented an improved event-plane method for
the flow analysis, which automatically corrects for autocorrelations and
nonflow effects. 
As in the standard method, each event has its 
{\it event plane\/} $\Phi_{EP}$, an estimate of the reaction plane, which
is the same as for the standard method, except for technical details
in the practical implementation. The trick which removes
autocorrelations and nonflow effects is that there is in addition an
{\it event weight\/}.  Anisotropic flow $v_n$ is then estimated as a {\it
weighted\/} average of $\cos(n(\phi-\Phi_{EP}))$. 
A straightforward application of this method would be to measure jet
production with respect to the reaction plane at LHC. With the
traditional event-plane 
method,  such a measurement would require to subtract particles
belonging to the jet from the event plane; in addition, strong nonflow
correlations are expected within a jet, which would bias the
analysis.

\section*{Acknowledgments}
JYO thanks Yiota Foka for a discussion which motivated this work. 
We thank S. Voloshin for comments on the manuscript.
The work of AB, NvdK and RS was supported in part by 
the Dutch funding agencies FOM and NWO.

\appendix
\section{Practical implementation}
\label{s:implementation}

Before we describe the implementation of the method, let us mention
that there are in fact two Lee-Yang-zeroes methods, depending on how 
the generating function is defined: the ``sum generating function'' 
makes explicit use of the flow vector~\cite{Bhalerao:2003yq}, while
the ``product generating function''~\cite{Borghini:2004ke}
is constructed using the azimuthal angles of individual particles, 
and cannot be expressed simply in terms of the flow vector. 
Cumulants also exist in both versions, the 
``sum''~\cite{Borghini:2000sa} and the
``product''~\cite{Borghini:2001vi}. 
For Lee-Yang zeroes, both the sum and the product give essentially the
same result for the lowest harmonic~\cite{Bastid:2005ct}: 
the difference between results from the two methods is
significantly smaller than the statistical error. 
On the other hand, the product generating function is significantly 
better than the sum generating function if one analyzes $v_4$ or 
$v_1$~\cite{Borghini:2004ad}  using mixed harmonics. 
The method described below is strictly equivalent to the sum 
generating function, although expressed in different terms. 
On the other hand, the product generating function cannot be recast in a 
form similar to the event-plane method, and will not be used here.

The method a priori requires two passes through the data, which are 
described in Sec.~\ref{s:findingzeroes} and Sec.~\ref{s:wandpsi}.

\subsection{First pass: locating the zeroes} 
\label{s:findingzeroes}

As with other flow analyses, one must first select events in some
centrality class. The whole procedure described below must 
be carried out independently for each centrality class.

The flow vector $(Q_x,Q_y)$ is defined by Eq.~(\ref{flowvector}).
In contrast to the standard event-plane
method, no flattening procedure is required to make the distribution 
of ${\bf Q}$ isotropic. Corrections for azimuthal anisotropies in 
the acceptance, which do not vary significantly in the event sample
used,
are handled using the procedure described below. 
We do not define the event plane $\Phi_{EP}$ as the 
azimuthal angle of the flow vector, as in Eq.~(\ref{flowvector}).
The procedure below defines both the event weight and the event plane. 

The analysis uses the projection of the flow vector onto an arbitrary
direction, see Eq.~(\ref{defqtheta}). 
In practice, the first pass should be repeated for 
several equally-spaced values of $n\theta$ between 
$0$ and $\pi$. 
This reduces the statistical error as shown by Eq.~(\ref{statvprime}).  
For more than 5 values of $\theta$ the reduction is not significant anymore, so this number is recommended. 
For elliptic flow, for instance, $\theta$ takes the values  
$\theta=0,\pi/10,2\pi/10,3\pi/10,4\pi/10$. 

One first computes the modulus $|G_{\theta}(r)|$, with $G_\theta$
defined by Eq.~(\ref{genx}), as a function of $r$ for positive
$r$. One determines numerically the first minimum of this
function. This is the Lee-Yang zero.  We denote its value by
$r_\theta$. It must be stored for each $\theta$.  

\subsection{Second pass: determining the event weight, $w_R$, 
and the event plane, $\Phi_{EP}$.}
\label{s:wandpsi}

In the second pass, one computes and stores, for each $\theta$, the following 
complex number:
\begin{equation}
\label{defdenom}
D_{\theta}\equiv \frac{1}{j_{01}N_{\rm evts}}\sum_{\rm events} r_\theta
Q_\theta e^{i r_\theta Q_\theta},
\end{equation}
where $j_{01}\simeq 2.40483$.
Except for statistical fluctuations and asymmetries in the detector
acceptance, $D_\theta$ should be purely imaginary. 

For each event, the event weight and the event plane are defined by 
\begin{eqnarray}
\label{wandpsi}
W_R\cos n\Phi_{EP}&\equiv& \left\langle {\rm Re}\left(\frac{e^{i r_\theta
 Q_\theta}}{D_{\theta}}\right)\cos n\theta\right\rangle_{\theta} \cr
W_R\sin n\Phi_{EP}&\equiv& \left\langle {\rm Re}\left(\frac{e^{i r_\theta
 Q_\theta}}{D_{\theta}}\right)\sin n\theta\right\rangle_{\theta},
\end{eqnarray}
where Re denotes the real part, and angular brackets denote averages
over the values of $\theta$ defined in subsection
\ref{s:findingzeroes}. 
Our estimate of $v_n$, denoted by 
$v_n\{{\rm LYZ}\}$, is then defined by Eq.~(\ref{v2LYZ}).

\begin{figure}
\centerline{\includegraphics*[width=1.\linewidth]{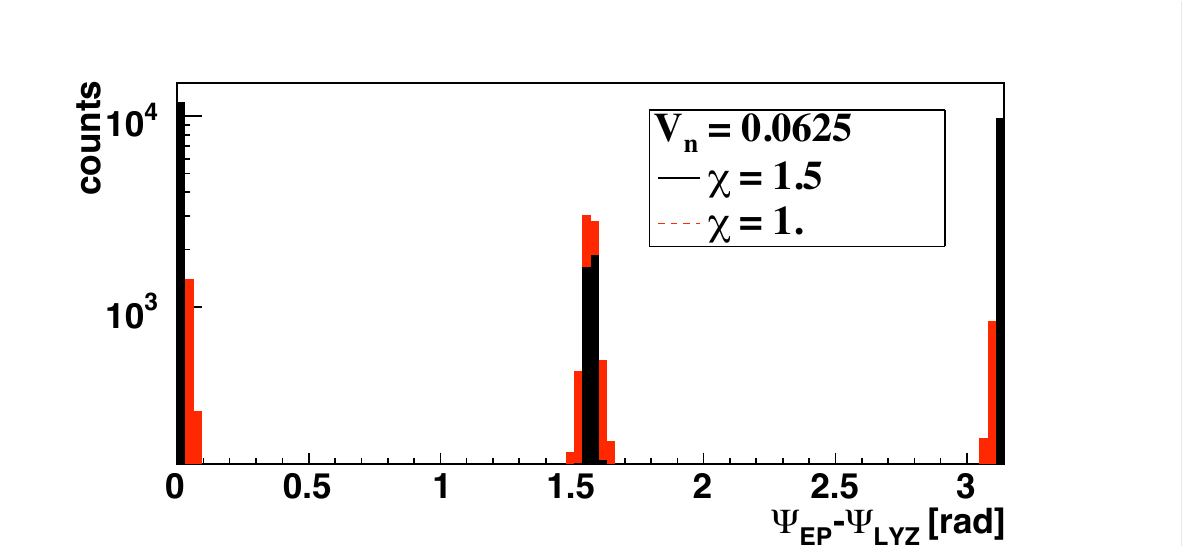}}
\caption{(Color online)
Distribution of the relative angle between the event plane
$\Phi_{EP}$ defined by Eq.~(\ref{wandpsi}), with $W_R>0$, and the 
standard event plane, for the reconstruction shown in
Fig.~\ref{fig:simulation}.} 
\label{fig:LYZvsstd}
\end{figure}

We now discuss how the angle $\Phi_{EP}$ defined by Eq.~(\ref{wandpsi})
compares with the event-plane from the standard analysis. 
First, we note that Eqs.~(\ref{wandpsi}) uniquely determine the angle 
$n\Phi_{EP}$ (modulo $2\pi$) only if the sign of $W_R$ is known. 
The simplest convention is $W_R>0$. 
In the simplified implementation described in Sec.~\ref{s:description}, 
however, where $\Phi_{EP}$ coincides with the standard event plane,
$W_R$ defined by Eq.~(\ref{wsimple}) can be negative, because the 
Bessel function changes sign (see Fig.~\ref{fig:weight}).
The convention $W_R>0$ then leads to a value of $n\Phi_{EP}$ 
which differs from the standard event plane by $\pi$, since 
changing the sign of $W_R$ amounts to shifting $n\Phi_{EP}$ by $\pi$ in 
Eqs.~(\ref{wandpsi}). 
This is illustrated in Fig.~\ref{fig:LYZvsstd}, which shows the distribution of the 
relative angle between $\Phi_{EP}$ and the standard event plane in the 
simulation of $v_2$ at LHC described in Sec.~\ref{s:results}. 
The distribution has two sharp peaks at 0 and $\pi/2$. The sign 
ambiguity produces the peak at $\pi/2$. The width of the peaks
results from statistical fluctuations. 
The final result $v_n\{{\rm LYZ}\}$, given by Eq.~(\ref{v2LYZ}), 
does not depend on the sign chosen for $W_R$. 

If one wishes to have an event-plane as close as possible to the standard 
event plane, one may choose the following convention.
Denoting by $\Phi_{EP}^{\rm std}$ the standard event plane, one computes
the following quantity:
\begin{equation}
S\equiv W_R\cos n\Phi_{EP} \cos n\Phi_{EP}^{\rm std}+
W_R\sin n\Phi_{EP} \sin n\Phi_{EP}^{\rm std},
\end{equation}
where $ W_R\cos n\Phi_{EP} $ and  $W_R\sin n\Phi_{EP}$ are defined by
Eq.~(\ref{wandpsi}). The sign of $W_R$ is then chosen as the sign of
$S$, which ensures that $n\Phi_{EP}-n\Phi_{EP}^{\rm std}$ lies between 
$-\pi/2$ and $\pi/2$.

The procedure described in this Appendix differs from the procedure
described in Sec.~\ref{s:description} only in the case of non-uniform
acceptance. 
This agreement can be seen in Fig.~\ref{fig:weight}, which displays a
comparison between the two. 
The solid line corresponds to the weight defined in 
Sec.~\ref{s:description} (Eqs.~(\ref{wsimple}) and (\ref{defC})),
while the stars corresponds to the weight defined by
Eq.~(\ref{wandpsi}), as implemented in the Monte-Carlo simulation
presented in Sec.~\ref{s:results}. The agreement is very good. 
This agreement can also be seen directly from the equations. 
If the detector has perfect azimuthal symmetry, $r_\theta$
and $D_\theta$ in Eq.~(\ref{wandpsi}) are independent of $\theta$, up
to statistical fluctuations. Neglecting these fluctuations, replacing
$Q_\theta$ with Eq.~(\ref{defqtheta}) and integrating over $\theta$,
one easily recovers Eq.~(\ref{wsimple}). 
If there are azimuthal asymmetries in the detector acceptance, on the
other hand, they are automatically taken care of by
Eq.~(\ref{wandpsi}). 
The fact that one first projects the flow vector onto a fixed
direction $\theta$ is essential 
(for a related discussion, see \cite{Selyuzhenkov:2007zi}).

\subsection{Statistical errors}
\label{s:stat}

The statistical error strongly depends on the 
resolution parameter~\cite{Ollitrault:1997di} $\chi$, which is closely 
related to the reaction plane resolution in the event-plane analysis.  
It is given by 
\begin{equation}
\label{defchi}
\chi=\frac{V_n}{\sqrt{\left\langle Q_x^2+Q_y^2\right\rangle-\left\langle Q_x\right\rangle^2 
-\left\langle Q_y\right\rangle^2-V_n^2}}.
\end{equation}
In this equation, $V_n$ is given by Eq.~(\ref{vnzero}), averaged over $\theta$ to 
minimize the statistical dispersion. The average values 
$\left\langle Q_x\right\rangle$, $\left\langle Q_y\right\rangle$,
$\left\langle Q_x^2\right\rangle$ and $\left\langle Q_y^2\right\rangle$ 
must be computed in the first pass through the data. 
Note that $\left\langle Q_x\right\rangle$ and $\left\langle Q_y\right\rangle$
vanish for a symmetric detector: they are acceptance corrections. 

The price to pay for the elimination of nonflow effects is an increased 
statistical error.
This increase is very modest if $\chi$ is larger than 1: If $\chi = 1.5$,
the error is only $25\%$ larger than with the standard event-plane method.
If $\chi=1$, it is larger by a factor 2. If $\chi=0.6$, it is 20 times larger. 
This prevents the application of Lee-Yang zeroes
in practice for $\chi$ smaller than 0.6. 

We now recall the formulas~\cite{Bhalerao:2003xf} which determine the
statistical error $\delta v_n^{\rm stat}$ on $v_n\{{\rm LYZ}\}$:
\begin{eqnarray}
\label{statvprime}
\lefteqn{(\delta v_{n}^{\rm stat})^2  = \frac{1}{4N^\prime J_1(j_{01})^2p}\sum_{k=0}^{p-1} \cos\left(\frac{k\pi}{p}\right)}& &\cr
& & \times\left[
\exp\!\left(\!\frac{j_{01}^2}{2\chi^2} \cos\left(\!\frac{k\pi}{p}\!\right)\!\right)
J_0\!\left(\! 2j_{01}\sin\!\left(\!\frac{ k\pi}{2p}\!\right)
\!\right)\right.\cr
& & -\left.\exp\!\left(-\frac{j_{01}^2}{2\chi^2} \cos \left(\!\frac{k\pi}{p}\!\right)\!\right)
J_0\!\left(\! 2j_{01}\cos\!\left(\!\frac{k\pi }{2p}\!\right)
\!\right) \right]
\end{eqnarray}
where $N^\prime$ denotes the number of  of objects one correlates to 
the event plane,  whatever they are (jets, individual particles),
and $p$ is the number of equally-spaced values of $\theta$ used in 
the analysis (see above). The larger $p$, the smaller the error. 
The recommended value is $p=5$, because larger values do not
significantly reduce the error. 
This equation shows that the statistical error diverges exponentially 
when $\chi$ is small.


\begin{thebibliography}{99}
\bibitem{footnote1}
In the event-plane method, the sum runs over a selected subset of 
detected particles, not over all particles, in order to suppress nonflow correlations. 
With the present method, nonflow correlations are not an issue. 
On the other hand, the method can only be applied if the resolution is large enough, 
as will be discussed in Sec.~\ref{s:discussion}. 
In order to maximize the resolution, one must use all detected particles.
\bibitem{footnote2}
$\chi$ depends on the average elliptic
flow $\langle v_2\rangle$, and on the event  multiplicity $M$. If
there are no nonflow effects, and if the weights in
Eq.~(\ref{flowvector}) are identical for all particles,
$\chi=\langle v_2\rangle \sqrt{M}$.
\end{thebibliography}
\end{document}